\documentclass[twocolumn,showpacs,showkeys,preprintnumbers,amsmath,amssymb]{revtex4}

\usepackage{graphicx}  
\usepackage{dcolumn}   
\usepackage{bm}        

\usepackage{amssymb}
\usepackage{amsfonts}
\usepackage{latexsym}

\def\beq{\begin{eqnarray}}
\def\eeq{\end{eqnarray}}
\def\ln{\,\mbox{ln}\,}

\def\Tr{\,\mbox{Tr}\,}

\def\al{\alpha}
\def\be{\beta}

\def\ga{\gamma}\def\de{\delta}

\def\ep{\epsilon}

\def\ka{\kappa}
\def\la{\lambda}
\def\na{\nabla}
\def\pa{\partial}

\def\si{\sigma}

\def\ph{\varphi}

\def\Ga{\Gamma}
\def\De{\Delta}
\def\La{\Lambda}

\begin{document}

%

\preprint{hep-th/0609138}

\title{Renormalization Ambiguities and Conformal Anomaly
in  Metric-Scalar Backgrounds}

\author{M. Asorey}
\affiliation{Departamento de F\'{\i}sica
Teorica, Universidad de Zaragoza, 50009, Zaragoza, Spain}
\author{G. de Berredo-Peixoto}
\email{guilherme@fisica.ufjf.br}
\author{I. L. Shapiro}
\altaffiliation{On leave from Tomsk State
Pedagogical University, Russia}
\email{shapiro@fisica.ufjf.br}
\affiliation{Departamento de F\'{\i}sica -- ICE,
Universidade Federal de Juiz de Fora,
Juiz de Fora, 36036-330, MG,  Brazil}

\begin{abstract}
We analyze the problem of the existing ambiguities in the
conformal anomaly in theories with external scalar field in curved
backgrounds. In particular, we consider the anomaly of
self-interacting massive scalar field theory and of Yukawa model
in the massless conformal limit. In all cases the ambiguities are
related to finite renormalizations of a local non-minimal terms in
the effective action. We point out the generic nature of this
phenomenon and provide a general method to identify the theories
where such an ambiguity can arise.
\end{abstract}

\pacs{04.62.+v, 11.10.Hi, 11.10.Jj}

\keywords{Curved space-time,  Non-Minimal
Interaction,  Renormalization Group,  Conformal Anomaly}

\maketitle


\section{\label{1}Introduction}

Fundamental scalar fields constitute an important element of the
standard model of particle physics and modern cosmology, in spite
that there is not yet any experimental confirmation of their
existence. Indeed, the cosmological applications of the scalar
fields (e.g. inflaton \cite{inflaton}, cosmon \cite{cosmon} and
quintessence \cite{quint}) require formulating them on curved
backgrounds at both classical and quantum level. It is well known
that the consistent  description of scalar field on curved
backgrounds  (see, e.g. \cite{book} for an introduction) is based
on the non-minimal covariant action \ $S = S_{scal}+S_{vac}$,
where
\beq
S_{\mathrm scal} &=& \frac12\int d^4x
\sqrt{g}\Big\{ g^{\mu\nu}\pa_\mu \phi\pa_\nu \phi
               + m^2\phi^2 + \xi R\phi^2
\label{scalar action}
\\
&-& \frac{\la}{12}\phi^4
+ \tau_1 \Box \phi^2
-\frac{\tau_2}{2}\,m\phi^3 - \ka\, m^2 \phi - \rho\,m\phi R
 \Big\}\,,
\nonumber
\eeq
and
\beq
S_{vac}\, = \,\int
d^4x\,\sqrt{g}\,\Big\{\,-\, \frac{1}{16\pi G}\,\left(R +
2\La\right) \,+\,a_1 C^2
\nonumber
\\
+ a_2 R^2 + a_3 E + a_4
\Box R \,\Big\}\,.
\label{vacuum}
\eeq
Here \ $C^2=R_{\mu\nu\al\be}^2
- 2R_{\al\be}^2 + 1/3\,R^2$ \ is the square of the Weyl tensor and
\ $E = R_{\mu\nu\al\be}^2-4 \,R_{\al\be}^2 + R^2$ \ is the
integrand of the Gauss-Bonnet topological invariant (Euler
number). $\,a_1,...,a_4$ and $\,G\,,\,\La\,$ are independent
parameters of the vacuum action,
$\,\la,\,\xi,\,\tau_1,\,\tau_2,\,\ka,\,\rho\,$ are independent
parameters in the scalar sector, including the non-minimal
parameters $\,(\xi,\,\rho)\,$ of interaction with the scalar
curvature
\footnote{We use the Euclidean metric $\eta_{\mu\nu}=\mbox{diag}(++++)$
and the definition
$\,R_{\mu\nu}=\partial_\lambda\,\Ga^\la_{\mu\nu}
 -\partial_\nu\,\Ga^\la_{\mu\la}+
\,\Ga^\sigma_{\mu\nu}\,\Ga^\la_{\sigma\la}-
\,\Ga^\sigma_{\mu\la}\,\Ga^\la_{\sigma\nu}$. We avoid using letter
$G$ for the Gauss-Bonnet invariant because it may be confused with
the Newton constant.}. The action $S$ contains  all local
diffeomorphism invariants up to  dimension four to guarantee the
renormalizability of the theory at quantum level. Terms with odd
powers of \ $\phi$ \ have been included to consider  models  with
spontaneous breaking of the \ $\phi\leftrightarrow-\phi$ \
symmetry \footnote{Later on, in section 4, we shall see that in
some situations these terms are indeed necessary.}.

The presence of  non-minimal terms like $R\phi^2$ and $R\phi$ and
$S_{vac}$ might be required because they may emerge as infinite
quantum corrections. In this case one needs such terms in the
classical action for renormalization purposes. These are not,
however, the only possible types of quantum loops contributions.
Some other important quantum corrections to $S$ of non-local type
might also appear. In the absence of scalar fields, for the
massless conformal case, these contributions can be partially
evaluated by means of the conformal anomaly (see, e.g.
\cite{birdav,duff77,AGS,brwcol} for some works on conformal
anomaly and \cite{rei,frts} for the anomaly-induced effective
action). In the case of massive quantum fields there is no general
method of calculation and one has to rely on some expansions which
are valid, typically, either for  large or small masses. In any
case the effective action is a functional of the metric and the
background scalar field, and the quantum corrections are somehow
related to the renormalization of the parameters $\,a_i$, $\,G$,
$\,\La\,$.

There are, however,  more involved situations, where the scalar
field is present at the cosmic scale. In this case the quantum
fields are interacting not only to the external metric but also
(directly or via the metric) to the external scalar field.
Therefore, in order to provide renormalizability, the vacuum
effective action has to include all local diffeomorphism
invariants of dimension four constructed from both metric and
scalar fields. In other words the whole scalar action \
$S_{\mathrm scal}$ \ must be considered as a part of the vacuum
action. In this situation it is important to extend the available
information concerning conformal anomaly and corresponding
derivation of the effective action to the theory with an external
scalar. It is known that for vanishing external scalar fields  the
anomaly manifests an ambiguity in the $\Box R$-sector
\cite{birdav,duff94}. In the recent paper \cite{AGS} we have
investigated this ambiguity in details. In particular, we have
shown that there is no conflict between the results for the
coefficient of the \ $\Box R$ term in the dimensional and other
regularizations, for the dimensional regularization leaves this
coefficient completely arbitrary. Furthermore, we have constructed
another example of covariant Pauli-Villars regularization with
similar ambiguity. In all cases the coefficient of the \ $\Box R$
term can be fixed by the renormalization condition for the finite
\ $\int R^2$-term in the vacuum action. Thus the ambiguity
concerns only the initial point of the renormalization group
trajectory for the corresponding parameter but not the shape of
this trajectory.

The purpose of the present paper is to extend the results of
\cite{AGS} for the case of a scalar field. We study the anomaly
and the effective action in the presence of a non-vanishing scalar
background and in particular investigate if new ambiguities appear
in that case. Indeed, the scalar may also interact with other,
e.g. fermion fields. Therefore, a complete understanding of the
problem requires to consider  also the quantum effects of these
interactions. Finally, we consider two distinct models:
self-interacting scalar field theory and Yukawa model for the
scalar-spinor interactions. In  both cases we shall find the
anomaly via the massless conformal limit in the effective action
for massive fields, not only by using dimensional regularization
\cite{duff77} (see also \cite{AGS}), but also by  covariant
Pauli-Villars regularization. The use of this regularization has
proved very fruitful in the analysis of the anomalous ambiguity in
the case of pure gravitational background  \cite{AGS} and we shall
extend these results  for the case with non-vanishing scalar field
backgrounds.

The paper is organized as follows. In next section we formulate
the conformal symmetry and anomaly for the scalar backgrounds in
curved space-times. The derivation of anomaly is performed by
means of dimensional regularization \cite{duff77} and the
discussion of corresponding ambiguities will parallel that of Ref.
\cite{AGS}.

In sections 3 and 4 we analyze the divergence structure in the
self-interacting scalar theory and in the Yukawa model.
Furthermore, we use Barvinsky-Vilkovisky formalism
\cite{bavi90,Avramidi} and the approach developed in \cite{apco}
to derive the non-local finite part of the effective action for
the case of massive Yukawa model. In the case of self-interacting
scalar theory, this calculation has been already done in
\cite{bexi}. The results for massive theories enable us to apply
the covariant Pauli-Villars regularization for the comprehensive
analysis of the nature of the new ambiguities due to the presence
of scalar background. Finally, in section 5 we outline our
conclusions concerning the general status of local conformal
symmetry at quantum level.

\section{\label{2} Local conformal symmetry and dimensional
regularization}

It is well known that the global conformal symmetry $SO(4,2)$ is
usually broken by quantum radiative corrections. At the same time
there is another similar symmetry associated to the background
fields. This symmetry is local\footnote{The relation between local
and global conformal symmetries have been discussed, in
particular, in \cite{raifear}.} and is not directly related to
space-time transformations, however it is also broken by anomaly.
The local conformal symmetry corresponds, in the case of the
scalar field $\phi$, to the transformation law \beq \phi \to
\phi^\prime = \phi \cdot e^{-\si}\,,\quad g_{\mu\nu} \to
g_{\mu\nu}^\prime = g_{\mu\nu} \cdot e^{2\si} \label{conformal}
\eeq where $\si = \si(x)\,$ is an space-time scalar function. This
local conformal symmetry reduces to Weyl symmetry when there is no
scalar background fields $\phi$. The action $S$ is invariant under
this symmetry if the following conditions are satisfied: \beq m=0
\,,\quad \xi=\frac16 \,,\quad \frac{1}{G}=0\,,\quad a_2=0\,.
\label{conditions} \eeq The corresponding Noether identity is \beq
{\cal T} \,=\, - \frac{2}{\sqrt{g}}\,g_{\mu\nu}\,\frac{\de S}{\de
g_{\mu\nu}} + \frac{1}{\sqrt{g}}\,\phi\,\frac{\de S}{\de
\phi}\,=\,0\,. \label{Noether} \eeq The first term is the usual
trace of the Energy-Momentum tensor while the second term is a new
contribution due to the presence of a background scalar. If the
scalar satisfies the classical equation of motion the second term
vanishes and the Noether identity  (\ref{Noether}) reduces to the
standard condition for Weyl invariance.

The existence of a quantum anomaly implies the violation of the
identity (\ref{Noether}). In absence of scalar background fields,
the simplest way to discover the conformal anomaly is via
dimensional regularization \cite{duff77}. However, the derivation
of the anomaly in dimensional regularization may exhibit some
ambiguities \cite{AGS} and these ambiguities might become even
larger in the presence of scalar background fields.

Dimensional regularization proceeds by  extending the covariant
perturbative expansion to an arbitrary dimension $n$. This is
achieved by replacing the differentials and integrals in the
action (\ref{scalar action}) by the $n$-dimensional covariant
derivatives and integrals, e.g.
\beq
\frac{1}{\sqrt{g}}\sum_{\mu,\nu=1}^4 \pa_\mu g^{\mu\nu}
\sqrt{g}\,\pa_\nu &\to& \frac{1}{\sqrt{g}}\sum_{\mu,\nu=1}^n \pa_\mu
g^{\mu\nu} \sqrt{g}\,\pa_\nu ;
\nonumber
\\
\int d^4 x &\to& \int d^nx\,.
\label{n}
\eeq
The  dimensionality of the different terms in the
action for $n\neq 4$ is restored by inserting an appropriate power
of the dimensionfull parameter $\mu$ which sets the
renormalization scale.

The first observation is that the purely metric part of the action
(\ref{vacuum}) is not conformal invariant in \ $n$ \ dimensions
independent on the values of the parameters of the action, in both
cases of global and local transformations. The situation in the
scalar sector (\ref{scalar action}) is more complicated. In fact,
the procedure (\ref{n}) leaves the room for different
prescriptions to extend the action of the scalar field. In order
to satisfy the symmetry under the global conformal symmetry
(\ref{conformal n}) one needs to comply with the same conditions
(\ref{conditions}). However, in order to provide the local
conformal symmetry one has to generalize the transformation law
for the scalar field and the value of the non-minimal parameter \
$\xi$, correspondingly, to \beq \phi \to \phi^\prime = \phi
\cdot\exp\Big( \frac{2-n}{2}\,\si \Big) \,,\qquad
\xi(n)=\frac{n-2}{4(n-1)}\,. \label{conformal n} \eeq An important
observation is that, in the classical theory, the introduction of
conformal coupling is not really necessary. After the removal of
the regularization by taking \ $n\to4$ limit, the theory is the
same independently on whether we follow (\ref{conformal n}) or
leave the transformation for scalar in the original form
(\ref{conformal}) and keep \ $\xi=1/6$ \ in \ $n$ dimensions.

Now let us discuss the violation of conformal symmetry on quantum
level in the framework of dimensional regularization. There are
two alternative, although equivalent, ways of deriving the
anomaly. Let us start from the one which is close to the original
derivation of Duff \cite{duff77} and is based on the direct use of
the counterterms. The second method is based on the evaluation of
the finite part of the effective action \cite{bavi-anom}, which
becomes anomalous after subtracting divergencies. This method will
be used in next sections to check the correctness of the
derivation based on the counterterms.

It proves useful to rewrite the Noether identity (\ref{Noether})
in different field variables. Let us define \ $g_{\mu\nu}= {\bar
g}_{\mu\nu}\cdot e^{2\si}$ \  and \ $\phi= {\bar \phi}\cdot
e^{-\si}$, where ${\bar g}_{\mu\nu}$ \  and \ ${\bar \phi}$ \ are
some fiducial fields. It is important to keep the original
four-dimensional form of the transformation law (\ref{conformal})
for the scalar, instead of the generalized one (\ref{conformal
n}). In the new variables the conformal transformation is reduced
to the shift of $\si$, while (\ref{Noether}) is cast in the new
form
\beq
{\cal T} =\left. - \frac{1}{\sqrt{{\bar
g}}}\frac{\de}{\de \si} S\left(n;{\bar g}_{\mu\nu}
e^{2\si}, {\bar \phi}e^{\frac{2-n}{2}\si}\right)
\right|_{\tiny
{\begin{array}{c}
\si\to 0  \\
 n\to 4      \end{array}}}
\,=\,0\,,
\label{conditions sigma}
\eeq
where the procedure includes also the replacements
${\bar g}_{\mu\nu}\to g_{\mu\nu}\,,\,\,{\bar \phi}\to \phi$.

At quantum level, the effective action which includes classical
action plus the naive one-loop corrections $\,\Gamma = S +
\bar{\Gamma}^{(1)}\,$ is locally conformal invariant at $n$
space-time dimension. Now, the presence of the UV divergences in
$\, \bar{\Gamma}^{(1)}\,$ requires introducing local counterterms,
${\Delta S}^{(1)}$, which explicitly break the conformal symmetry.
In the vicinity of $n=4$ the requested counterterms are defined as
\beq
{\Delta S}^{(1)}\!\left(n,g_{\mu\nu},\phi\right)
&=&\frac{1}{n-4}\int d^4x \sqrt{g}\Big\{
\beta_1C^2\!+\!\beta_2E\!+\!\beta_3 \Box R
\nonumber
\\
&+&   \al_1R\phi^2
\!+\! \al_2\Box\phi^2 \!+\! \al_3\phi^4 \Big\}\,,
\label{divs}
\eeq
$\be_i$ and $\al_i$  being the $\be$-functions in the vacuum  and
scalar matter sectors, respectively. The effect of these terms
into conformal Noether identity is
\beq
<{\cal T}> =\left. - \frac{1}{ \sqrt{\bar g} }
\, \frac{\delta \De S^{(1)}\left(n;\,{ g}_{\mu\nu} e^{2\si},\,
\phi e^{-\,\si}\right)}{\de \si}\,
 \,\right|_{\tiny
{\begin{array}{c}
\si\to 0  \\
 n\to 4      \end{array}}}
\nonumber
\\
=\be_1 C^2 + \be_2 E + \be_3 \Box R + \al_1 R\phi^2
+\al_2\Box\phi^2 + \al_3\phi^4.
\label{ann}
\eeq
Let us remark that this expression is exactly the
anomaly of the global conformal symmetry and hence it can be
obtained via the renormalization group \cite{tmf,book} or the \
$\zeta$-regularization method \cite{haw}. The expression for the
conformal anomaly (\ref{ann}) has been used to get an explicit
form of the anomaly-induced action in the scalar-metric theory
(cosmon model) \cite{asta}, with very interesting phenomenological
implications. In next sections we will show that the approach
described above agrees with the calculation based on the
heat-kernel \cite{bavi90} and Pauli-Villars methods.

Before discussing the source of ambiguities in the dimensional
regularization, let us make some general remarks. The counterterms
in $\De S$ are local expressions on the background fields, whereas
among the finite terms of the effective action
$\bar{\Gamma}^{(1)}$ there are both local and non-local terms. The
universality of non-local terms \cite{univamb,jll} guarantees the
universality of the terms in the anomaly which are not total
derivatives. In the particular case of metric and scalar
background fields this means that the ambiguities should only
affect the coefficients $\alpha_2$ and $\be_4$ of the terms
$\Box\phi^2$ and $\Box R$ and the corresponding local terms
$R\phi^2$ and $R^2$ in the finite part of the effective action.
The ambiguities arise from the conformally invariant local terms
in the counterterms $\De S^{(1)}$ which are total derivatives or,
equivalently, from the finite local anomalous terms of
$\bar{\Gamma}^{(1)}$. Qualitatively the reason for this is as
follows. The splitting of the  quantum effective action into
divergent and finite parts is ambiguous and the actual expression
for the conformal anomaly becomes dependent on this splitting in
the local terms which break local conformal symmetry. In
principle, the ambiguity has to be fixed by the choice of a
suitable renormalization scheme \cite{birdav,AGS}. However, as we
shall see below, in the presence of background scalar this may be
a difficult task to accomplish.

The ambiguity in the splitting can be easily understood within the
dimensional regularization scheme. The conditions for constructing
the counterterm of the $R\phi^2$ type are locality and
cancellation of the divergent part of the effective action.
However, both requirements  allow, instead of the minimal choice
$h_1=\frac12 \left[\xi-\frac{n-2}{4(n-1)}\right]$ in the
$\,R\phi^2$ term of (\ref{divs}), a more general choice
$$
h_1\,=\,\frac12\left[\xi-\frac{n-2}{4(n-1)}+\ka(n-4)\right]\,,
$$
where $\ka$ is an arbitrary dimensionless parameter. For any value
of $\ka$ the counterterm is local and it cancels the UV
divergence. Hence we meet an ambiguity. The corresponding term in
the anomaly (\ref{ann}) depends on $\ka$ \beq 6 {h_1}\,\Box\phi^2
\,\to\, 6 \left[{h_1}+ \ka \right]\,\Box\phi^2\,.
\label{modification} \eeq The arbitrariness here is quite similar
to the one discussed in \cite{AGS} for the \ $\Box R$ \ term. In
the last case the anomaly corresponds to the local $\int
d^4x\sqrt{g}R^2$ \ term in the effective action, and therefore the
arbitrariness in the anomaly corresponds to the freedom of adding
a local finite $\int d^4x\sqrt{g}R^2$ term to the classical vacuum
action. In the present case the anomalous contribution corresponds
to the finite term \ $\int d^4x\sqrt{g}R\phi^2$. However, there is
a very important difference. Adding a finite \ $\int
d^4x\sqrt{g}R^2$ \ term to the classical vacuum action  does not
break any symmetry of the theory in the sector of scalar quantum
fields, while adding $\int d^4x\sqrt{g}R\phi^2$  may break the
conformal symmetry of the quantum sector of the theory. The
situation with such term is indeed closer to the one with the
$\int d^4x\sqrt{g}R^2$ term in conformal quantum gravity
\cite{Weyl} rather than in the semiclassical theory.

From the previous discussions it is also straightforward to
understand what is the general structure of the ambiguous terms in
any dimension: $\Box \phi^2$ and $R$ in two dimensions and
furthermore
$$
\Box \phi^2\,,\quad \Box
\left(R_{\mu\nu\al\be}^2\right)^{\frac{n}{2}-1}\,,\quad \Box
\left(R_{\al\be}^2\right)^{\frac{n}{2}-1}\,,\quad \cdots \,\,\,
\Box R^{\frac{n}{2}-1}
$$
for any even dimension $n$. The corresponding local terms of the
finite part of the effective actions may have the form
$$
R\phi^2\,,\quad R
\left(R_{\mu\nu\al\be}^2\right)^{\frac{n}{2}-1}\,,\quad R
\left(R_{\al\be}^2\right)^{\frac{n}{2}-1}\,,\,\, \cdots \,\,\,
R^{\frac{n}{2}}.
$$
In the case of background gauge fields there are also extra terms
like the Chern-Simons term in 3-dimensions \cite{pr,univamb}.

Let us discuss, for completeness, another possible implementation
of local conformal symmetry in dimensional regularization which
does not lead to the same result. Following the method introduced
in \cite{duff77} (see also \cite{brwcol,hath}) one could use a
different  transformation law for the background scalar field
(\ref{conformal n}) and the corresponding value of $\xi(n)$. The
peculiarity of this prescription  is that $S_{\mathrm scal}$
becomes invariant under local conformal transformations in
arbitrary dimension and, of course, the same is true for the
corresponding counterterms. Thus, although the pure gravitational
part of the anomaly is obtained as in the minimal prescription
\cite{AGS}, there is no $\phi$ dependent part of the anomaly. The
final expression for the conformal anomaly is reduced to
\beq
<{\cal T'}> &=&-\left.\frac{1}{\sqrt{\bar g}} \frac{\delta \De
S^{(1)} \big(n;{ g}_{\mu\nu} e^{2\si}, \phi
e^{\frac{2-n}{2}\si}\big)}{\de\si} \right|_{\tiny
{\begin{array}{c}
\si\to 0  \\
 n\to 4      \end{array}}}
\nonumber
\\
&=&\be_1 C^2 + \be_2 R^2 + \be_3 E + \be_4 \Box R ,
\label{def anomaly}
\eeq
that essentially differs from the correct one
(\ref{ann}). There are missing terms which depend on the scalar
background field $\phi$. A more exotic renormalization scheme,
leading to the total uncontrollable ambiguity in the conformal
anomaly, is demonstrated in the Appendix.

\section{\label{3} Self-interacting scalar field}

The derivation of the effective action for self-interacting scalar
field theory (\ref{scalar action}) has  already been carried out
in dimensional regularization \cite{bexi} and we shall merely
analyze the limit relevant for conformal anomaly. For the sake of
simplicity we consider the special case with  vanishing values of
the coefficients \ $\tau,\,\ka,\,\rho\,$ of the odd-power terms in
the action (\ref{scalar action}).

The one-loop effective action is defined, within the background
field method, as \beq \bar{\Ga}^{(1)}\,=\,-\,\frac{1}{2}\,{\rm
Tr\, ln} \,\left[\, -\hat{1}\,\Box - \hat{1}\,m^2 - \hat{P} +
\frac{\hat{1}}{6}\,R \,\right]\,, \label{trln} \eeq where the hats
indicate operators acting in the space of the quantized fields.
For the case of the real scalar field $\,\phi$, $\,{\hat 1}\,$ is
the image of the delta function and $\,\hat{P} = -(\xi - 1/6)R +
\la\phi^2/2$. The divergent part of effective action can be easily
calculated using the standard Schwinger-DeWitt technique
\beq
{\bar \Ga}^{(1)\,div}_{scalar}
\,=\,-\frac{1}{(4\pi)^2(n-4)}\,\int
d^4x \,\sqrt{g}\, \Big\{ \frac{\la}{12}\,\Box\phi^2
\nonumber
\\
\,-\,\frac{\la}{2}\,\Big(\xi-\frac16\Big)\Box R
\,+\,\frac{1}{180}\,\Box R \,+\, \frac{1}{120}\,C^2
- \frac{1}{360}\,E
\nonumber
\\
-\,\frac{1}{2}\,\Big(\xi-\frac16\Big)^2\,R^2
\,+\,\frac{\la}{2}\,\Big(\xi-\frac16\Big)\,R\phi^2
\,+\,\frac{\la^2}{8}\,\phi^4 \Big\}\,.
\label{surface divs}
\eeq

Let us notice that all the integrals in this section are
four-dimensional, because we use dimensional regularization in the
minimal way discussed in the previous section. Alternatively one
can apply the proper-time cut-off regularization \cite{bavi90,AGS}
here, the result is of course the same. Later on we shall see that
if instead of dimensional regularization we use the covariant
Pauli-Villars regularization \cite{AGS} the logarithmic divergent
part of the action  will agree with (\ref{surface divs}).

The calculation of both divergent and finite part of the effective
action has been performed up to the second order in the curvatures
$\,(R_{\mu\nu\al\be},\,{\hat P})\,$ using Feynman diagrams on flat
backgrounds and also the heat kernel solution
\cite{Avramidi,bavi90}. The result is \cite{apco,bexi}
\beq
{\bar \Ga}^{(1)}_{\mathrm scalar}
&=&\frac{1}{2(4\pi)^2}\,\int d^4x \,\sqrt{g}\,
\Big\{\,\frac{m^4}{2}\cdot\Big(\frac{1}{\ep} +\frac32\Big)
\nonumber
\\
&+& \Big(\xi-\frac16\Big)\,m^2R\, \Big(\frac{1}{\ep}+1\Big)
\nonumber
\\
&+& \frac12\,C_{\mu\nu\al\be}
\,\Big[\frac{1}{60\,\ep}+k_W(a)\Big] C^{\mu\nu\al\be}
\nonumber
\\
&+&
R\,\Big[\,\frac{1}{2\ep}\,\Big(\xi-\frac16\Big)^2\, +
k_R(a)\,\Big]\,R
\nonumber
\\
&-&\frac{\la}{2\ep}\,m^2\phi^2
\,+\,\phi^2\Big[\frac{\la^2}{8\ep}+k_\la(a)\Big]\phi^2
\nonumber
\\
&+&
\phi^2\Big[
-\frac{\la}{2\ep}\Big(\xi-\frac16\Big)+k_\xi(a)\Big]\,R
\,\Big\}\,,
\label{final}
\eeq
where
$$
\frac{1}{\ep}=\frac{2}{4-n} +\ln \Big(\frac{4\pi \mu^2}{m^2}\Big)
- \ga\,,
$$
$\ga$ being the Euler number. The relevant form factors are
\cite{apco,bexi}
\beq
k_\la(a) &=& \frac{A\, \la^2}{4}\,,
\label{form scalar}
\\
k_\xi(a) &=& \la\left[\,\frac{A\,(a^2-4)}{12\,a^2}
\,-\,\frac{1}{36}\,-\,A\,\Big( \xi - \frac16\Big)\,\right]\,,
\nonumber
\eeq
\beq
k_W(a) &=& \frac{8A}{15\,a^4}
\,+\,\frac{2}{45\,a^2}\,+\,\frac{1}{150}\,,
\nonumber
\\
k_R(a) &=&
A\Big(\xi-\frac16\Big)^2-\frac{A}{6}\,\Big(\xi-\frac16\Big)
+\frac{2A}{3a^2}\,\Big(\xi-\frac16\Big)
\nonumber
\\
&-&
\frac{A}{18a^2} + \frac{A}{9a^4} + \frac{A}{144}+\frac{1}{108\,a^2}
- \frac{7}{2160}
\nonumber
\\
&+&
\frac{1}{18}\,\Big(\xi-\frac16\Big)\,,
\label{W}
\eeq
where we used the notations \cite{apco} \beq
A\,=\,1-\frac{1}{a}\ln\frac{1+a/2}{1-a/2}\,, \qquad a^2 =
\frac{4\Box}{\Box - 4m^2}\,. \label{Aa} \eeq An important
difference between the divergences (\ref{surface divs}) and the
divergent part of (\ref{final}) is that the first expression does
include surface terms while the second one does not. Indeed, one
can always restore these terms using anomaly.

The conformal limit corresponds to \ $m^2\to 0$ \ and \ $\xi\to
1/6$. The useful relations corresponding to the massless limit \
$m^2\to 0$ \ are as follows:
\beq
a\to 2\,,\quad A \sim
\frac{1}{2}\,\log\left|2-a\right|\to \infty \,,\quad
(a-2)A \to 0\,. \label{a 2 2}
\eeq
Applying the limit  (\ref{a 2 2}) to the
expression (\ref{final}) reduces the  non-conformal term to
\beq
{\bar \Ga}^{(1)}_{reg}
&=&  -\,\frac{1}{2(4\pi)^2}\,\int  d^4x
\,\sqrt{g}\,\Big\{ \frac{\la^2}{8}\phi^2\, \log
\left(\frac{\square}{4\pi\mu^2}\right)\phi^2
\nonumber
\\
&+& \frac{1}{120}\,C_{\mu\nu\al\be}\,\log
\left(\frac{\square}{ 4\pi \mu^2}  \right) C^{\mu\nu\al\be}
\nonumber
\\
&+& \frac{\la}{36}\phi^2\,R \,+\,\frac1{1080}\,R^2\ \,\Big\}.
\label{remains}
\eeq
It is easy to see (using the formulas of
previous section) that this result perfectly fits with the
divergence (\ref{surface divs}). The coincidence holds for both
metric and metric-scalar local terms. Hence, at this level, the
heat-kernel solution does not show any sign of ambiguity discussed
in section 2.

Let us now analyze the explicit emergence of these ambiguities in
the framework of Pauli-Villars regularization. In this
regularization method the classical action (\ref{scalar action})
is supplemented with extra Pauli-Villars fields whose interactions
with the physical field $\phi$ are given by
\beq
S_{\rm reg} &=&
\sum_{i=1}^N\int d^4 x\sqrt{-g}
\,\Big\{\,\frac12\,g^{\mu\nu}
\partial_\mu\,
\varphi_i\partial_\nu\varphi_i +\frac{\xi_i}{2}\,R\,\varphi_i^2
\nonumber
\\
&-&
\frac{m^2_i}{2}\, \varphi_i^2\, -\frac{\lambda}{2}
\phi^2\varphi_i^2 \Big\}\,.
\label{regaction}
\eeq
The physical
scalar field $\phi$ (also labeled below by $\varphi_0$) is
conformally coupled ($\xi=1/6$) and has bosonic statistics
($s_0=1$). The $N$ Pauli-Villars fields $\,$ $\varphi_i$ $\,$
($i=1,\dots,N$) $\,$ are massive $\, m_i=\mu_i M\neq 0$ and may
have degeneracy \ $s_i$. These fields can have either bosonic or
fermionic statistics. In the first case the degeneracy of the
field $s_i$ has to by multiplied by the factor of $1$ and in the
former case by the factor of \ $-2$. We also assume, for the sake
of completeness, that the Pauli-Villars regulators might have
non-conformal couplings $\xi_i\neq 1/6$. The regularized effective
action of the massless scalar field with conformal coupling  \
$\xi=1/6$ \ is given by \beq {\bar \Ga}^{(1)}_{\rm reg}\,=\,
\lim_{\Lambda\to \infty}\sum_{i=0}^N s_i {\bar \Ga}^{(1)}_{\rm i}
\left(m_i,\xi_i,\Lambda\right)\,, \label{total} \eeq where
$\Lambda$ is an auxiliary momentum cut-off.

According to the general prescription, all divergences in the
ultraviolet cut-off $\Lambda$ are cancelled out due to the
Pauli-Villars conditions
\beq
\sum_{i=1}^N \,{s_i}\,=\,-\,s_0\,=\,-\,1\,;
\label{23a}
\\
\sum_{i=1}^N \,s_i\,\mu_i^2\,=\, 0\,;
\sum_{i=1}^N \,s_i\,\left(\,\xi_i-\frac16\,\right)\,=\,0\,;
\label{23b}
\\
\sum_{i=1}^N\,s_i\, \mu_i^4\,=\, 0\,;
\sum_{i=1}^N\,s_i\,\left(\xi_i-\frac{1}{6}\right)^2\,=\,0\,.
\label{23c}
\eeq
 which are identical to those of the free scalar fields
\cite{AGS}. The first equation (\ref{23a}) cancels out quartic
divergences $\,\Lambda^4$, the second and third equations
(\ref{23b}) cancel quadratic ones $\,\Lambda^2\,$ and the last
two equations (\ref{23c}) are required to cancel logarithmic
divergences $\,\,\log\,(\Lambda^2/m^2)$. A simple solution of
these equations matching all these requirements is
\beq
&\displaystyle{s_1\,=\,1,\,\qquad s_2\,=\,4,\,\qquad s_3\,=-\,s_4
\,=\,\,s_5\,=-\,2}, \phantom{\Big ]} \cr
&\displaystyle{\mu_1^2\,=\,4,\,\quad \mu_2^2\,=\,3, \,\quad
\mu_3^2\,=\,1,\,\quad \mu_4^2\,=\,3,\,\quad \mu_5^2\,=\,4}
\phantom{\Big ]}
\nonumber
\\
\cr
& \quad \mbox{and} \qquad \xi_i=\mu^2_i+\frac16\,.
\label{solutions mu xi}
\eeq
Another possible solution emerges if one takes all \
$s_i,\,\mu_i$ \ as in (\ref{solutions mu xi}) and \ $\xi_i\equiv
1/6$. At that point we detect an ambiguity in the solution for the
regularized effective action.

A compact expression for the effective action is obtained in the
limit $\,M\to \infty\,$. In this limit the form factors
$\,k_W(a)\,$ and $\,k_R(a)\,$ for the auxiliary regulator fields \
$\varphi_{1,...,5}$ \ vanish and the asymptotic form of the
remaining expression has the form
\beq
{\bar \Ga}^{(1)}_{reg}
&=&
\displaystyle\,\frac{1}{2(4\pi)^2}\,\int d^4x
\,\sqrt{g}\,\Big\{\frac{M^4 \al}{2}
+ M^2 R\beta
\nonumber
\\
&+& \Big(\delta-\frac1{1080}\Big)R^2
\,-\,{\la}\Big(\frac{1}{36}+\sigma\Big)\phi^2\,R
\nonumber
\\
&-&  \frac{\la^2}{8}\phi^2\,
\log\Big(\frac{\square}{4\mathrm{e}^{\eta'} M^2}\Big)\phi^2
\nonumber
\\
&-&  \frac1{120}C_{\mu\nu\al\be}  \log
\Big(\frac{\square}{4\mathrm{e}^\eta M^2}\Big) C^{\mu\nu\al\be}
 \Big\},
\label{finall}
\eeq
where
\beq
\alpha=\sum_{i=1}^N s_i \mu_i^4\ln\,\mu_i^2\,,
\nonumber
\\
\beta=\sum_{i=1}^N s_i\, \mu_i^2\, \Big(\,\xi_i-\frac16\,\Big)\ln\,\mu_i^2\,,
\nonumber
\\
\eta=\frac{2}{3}-\sum_{i=1}^N
s_i\, \ln\, \mu_i^2\,=\frac{2}{3}+\eta'\,,
\nonumber
\\
\sigma=\frac12\sum_{i=1}^N s_i\, \,
\Big(\,\xi_i-\frac16\,\Big)\ln\, \mu_i^2\,,
\nonumber
\\
\delta =\sum_{i=1}^N s_i\,
\Big(\,\xi_i-\frac16\,\Big)^2\, \ln\, \mu_i^2\,,
\label{25}
\eeq
The finite part of the effective action (\ref{final}) which breaks
conformal invariance  has two kind of terms. First, the terms
which are non-local because of the presence of logarithmic in \
${\square}$ \ insertions and second those which are local. The
last terms essentially reduce to two types, $R\phi^2$ and $R^2$.
The first type of terms have universal coefficients because of
their non-locality. The second type of terms depend on arbitrary
parameters of the Pauli-Villars regularization \ $\delta$ \ and \
$\sigma$. These properties are inherited by their descendents in
the expression for conformal anomaly in the covariant
Pauli-Villars regularization
\beq
<{\cal T}>
&=&\frac{1}{(4\pi)^2}\,\Big[ \,\frac{1}{120}\,C^2 -
\frac{1}{360}\,E + \frac{\lambda^2}{8}\phi^4
\nonumber
\\
&+&
\Big(\frac{1}{180} - 6\de\Big)\,{\square}R
+ \Big( \frac{1}{12}
+ 3\sigma\Big)\square\phi^2\Big]\,.
\label{trace111}
\eeq
The last
expression shows that, whereas the coefficient of the
$\lambda\phi^4 $, Weyl and Euler terms in the anomaly are
universal, those of $\square R$ and $\square \phi^2$ terms are in
fact arbitrary, for they depend on the regularization procedure.
Qualitatively this is the same ambiguity we have detected in the
dimensional regularization case.

The analysis performed above can be easily expanded and we can
establish a general structure of the ambiguities. The ambiguous
terms of the effective action are always (up to the conformal
terms) of the form \ $R{\mathcal{A}}(R,\phi)$, where \
${\mathcal{A}}$ \ is a dimension \ $n-2$ \ local operator which
induce the anomaly terms of the type
$\square{\mathcal{A}}(R,\phi)$. Indeed, in  dimension $n=4$, we
have ${\mathcal{A}}=R$ or ${\mathcal{A}}=\phi^2$, in $n=2$,
${\mathcal{A}}={\mathrm{cte}}$ or ${\mathcal{A}}=\phi^k$, however
from the above consideration it is clear that in arbitrary
dimension \ $n$ \ the operator ${\mathcal{A}}=\phi^2$ is always
ambiguous.

\section{\label{4}The Yukawa model}

The Yukawa model in curved Euclidean space \beq S_{Yukawa} =
i\,\int d^4x \sqrt{g}\,\,{\bar \psi} \left(\,\ga^\mu\na_\mu - im -
ih\phi\,\right)\psi\,. \label{Yukawa} \eeq describes the
interaction of fermionic field \ $\psi$ \ with a scalar background
field $\phi$. Let us denote
\beq
\ph = m + h\phi\,,\quad \hat{H} =
\ga^\mu\na_\mu - i\ph\,. \label{H fermion}
\eeq
The  one-loop
effective action can be calculated in two different ways. The
first possibility is to consider
\beq
\bar{\Ga}^{(1)}_{fermion}\left[g_{\mu\nu},\phi\right] &=&
\Tr\ln\left(\hat{H}\cdot\hat{H}_1^{*}\right) -
\frac{1}{2}\Tr\ln\left(\hat{H}_1^{*}\cdot\hat{H}_1^{*}\right),
\nonumber
\\
\hat{H}_1^{*}&=&\ga^\nu\na_\nu
\label{one}
\eeq
and the
second corresponds to
\beq
\bar{\Ga}^{(1)}_{fermion}\left[g_{\mu\nu},\,\phi\right]
&=&
\frac12\,\Tr\ln\,\left(\hat{H}\cdot\hat{H}_2^{*}\right)
\nonumber
\\
\hat{H}_2^{*}&=&\ga^\nu\na_\nu + i\ph\,,
\label{two}
\eeq
where we
use the known fact \ $\bar{\Ga}^{(1)}_{fermion}$ \ is an even
functional in \ $\ph$. Let us notice that the derivation of the
second term in (\ref{one}) can be performed in a standard way
(see, e.g. the second reference in \cite{apco}).

In the first case (\ref{one}) the relevant  operator is
\beq
\hat{H}\cdot\hat{H}_1^{*}  &=&  \hat{1}\Box
+ 2\hat{h}^\al\na_\al + \hat{\Pi}\,,
\nonumber
\\
\hat{h}^\al &=&  -\frac{i\ph}{2}\,\ga^\al
\quad \mbox{and} \quad
\hat{\Pi}=-\frac{\hat{1}}{4}\,R\,.
\label{op one}
\eeq
The
calculation of divergences can be performed using standard
prescription \cite{hove,bavi81} (see also \cite{book}) and the
result is
\beq
{\bar \Ga}^{(1)\,div}_{fermion}(\hat{H}\cdot\hat{H}_1^{*})
&=& -\frac{1}{(4\pi)^2(n-4)}\,\int d^4x
\,\sqrt{g}\,\Big\{\,\frac{k}{2}\Box\ph^2
\nonumber
\\
&+& \frac{1}{30} \,\Box R
\,+\,2g^{\mu\nu}\pa_\mu\ph\pa_\nu\ph \,-\,2\ph^4
\nonumber
\\
&+&
\frac{1}{3}\,R\ph^2 \,+\, \frac{1}{20}\,C^2
\,-\, \frac{11}{360}\,E\Big\} \,.
\label{1f divs}
\eeq
where $k=k_1=-{8}/{3}$.

In the second approach  (\ref{two}) the relevant operator is
\beq
\hat{H}\cdot\hat{H}_2^{*}\,=\,\hat{1}\Box + \hat{\Pi}_1\,,
\quad \mbox{where}
\nonumber
\\
\hat{\Pi}_1=\ph^2+i\ga^\mu\ph_{,\mu}-\frac{\hat{1}}{4}\,R\,.
\label{op two}
\eeq
The divergences have the same form except that
the coefficient $k$ has a different value $k=k_2=-{4}/{3}$. The
difference in the coefficient $k$ is quite remarkable, for it
might indicate to the arbitrariness in the anomalous $R\phi^2$
term in the finite part of the effective action. Later on we shall
confirm that this is exactly the case.

The operator \ $\big(\hat{H}\cdot\hat{H}_1^{*}\big)$ \ in the
prescription (\ref{one}) has linear in derivative term and does
not admit direct application of the heat-kernel solution
\cite{bavi90}. The derivation of the full (with finite part)
effective action using the method of \cite{bavi90,apco} gives, in
the case of the prescription (\ref{two}), the following result:
\beq
{\bar \Ga}^{(1)}_{fermion}(\hat{H}\cdot\hat{H}_2^{*})
&=&
\frac{1}{(4\pi)^2}\,\int d^4x \,\sqrt{g}\,
\Big\{ -m^4\Big(\frac{1}{\ep}+\frac32\Big)
\nonumber
\\
&+&
\frac{1}{6}\,m^2R\,\Big(\frac{1}{\ep}+1\Big)
+ \frac12\,R\,k^f_R(a)\,R
\nonumber
\\
&-&
2m^2\,(\ph^2 - m^2)\,\Big(\frac{1}{\ep}+1\Big)
\label{1f finite}
\\
&+&
\frac14\,C_{\mu\nu\al\be}
\Big[\frac{1}{10\,\ep}+k^f_W(a)\Big] C^{\mu\nu\al\be}
\nonumber
\\
&+&
\,\frac{1}{2}\,\big(\na_\al\ph\big)\, \Big[\frac{2}{\ep} +
4A\Big]\,\big(\na^\al\ph\big)
\nonumber
\\
&+&
\frac12\,(\ph^2-m^2)
\Big[\frac{1}{3\ep}+\frac29+\frac{8A}{3a^2}\Big]\,R
\nonumber
\\
&-&
\frac12(\ph^2-m^2)\Big[\frac{2}{\ep}+4A\Big](\ph^2-m^2)
\,\Big\}\,.
\nonumber
\eeq
The higher derivative form factors $k^f_W$
and $k^f_R$ can be found in the second reference in \cite{apco}.
The change $\ph^2\to(\ph^2-m^2)=h^2\phi^2+2mh\phi$ is due to the
fact we have to settle $m^2$ into exponential of the heat kernel
representation for the effective action \cite{apco}.

The massless limit  (\ref{a 2 2}) of the finite part of the one
loop effective action (\ref{1f finite}) leads to the following
expression for the non-conformal term:
\beq
{\bar \Ga}^{(1)}_{reg}
&=& -\,\frac{1}{2(4\pi)^2}\,\int  d^4x \,\sqrt{g}\,\Big\{\,
\frac1{90}\,R^2 \,-\,\frac{2\,h^2}{9}\,\phi^2\,R
\nonumber
\\
&-&  2h^4\,\phi^2\,\log\left(\frac{\square}{ 4\pi\mu^2}\right)
\phi^2
\nonumber
\\
&+& \frac1{20}\,C_{\mu\nu\al\be}
\,\log \left(\frac{\square}{4 \pi \mu^2}  \right) C^{\mu\nu\al\be}
\nonumber
\\
&+&  2h^2\,\left(\na_\alpha\phi\right)
\,\log\left(\frac{\square}{
4\pi\mu^2}\right) \left(\na^\alpha\phi\right)
\nonumber
\\
&+& \frac{h^2}{3}\,\phi^2
\,\log\left(\frac{\square}{4\pi\mu^2}\right)\,R \,\Big\}.
\label{remains f}
\eeq
which is
quite similar to the result for the scalar case (\ref{remains}).
There are non-local finite terms which are in one to one
correspondence with the local UV divergent terms. These terms were
cancelled by counterterms in a momentum-subtraction
renormalization scheme (see \cite{apco} for a detailed
discussion). On the top of that, we have two usual local terms \
$\int\sqrt{g}R^2$ \ and \ $\int\sqrt{g}R\phi^2$. At this level no
ambiguity is observed.

More complete analysis can be performed in the framework of
Pauli-Villars regularization, in a way similar to that described
in the previous section. Our purpose is to detect the ambiguity in
the local terms \ $\int\sqrt{g}R^2$ \ and \ $\int\sqrt{g}R\phi^2$,
these ambiguities are expected to be similar to the ones we met
for the scalar case. In order to use this analogy we need to
introduce some parameters similar to the non-minimal ones \
$\xi_i$.

In order to get the most general ambiguity we shall introduce PV
regulator fields  \ $\psi_i$ with bilinear  couplings like in
(\ref{op two}) to include  non-minimal couplings
\beq
\left(\hat{H}\cdot\hat{H}_2^{*}\right)_{(i)} &=&
\hat{1}\Box + \hat{\Pi}_{1i}\,, \quad \mbox{where}
\nonumber
\\
\hat{\Pi}_{1i} &=&
\ph_i^2+i\ga^\mu\ph_{i,\mu}
+\Big(\chi_i-\frac{1}{4}\Big)\hat{1}R,
\label{op two i}
\eeq
where \ $\ph_i=h\phi+M_i$, \ $M_i=\mu_i\,M$ \ being the mass
of the regulator fields, $\,M\,$ is the regularization scale
parameter and \ $\mu_i$ \ dimensionless parameters which can be
defined from the condition of cancelling the divergences. Finally,
 \ $\chi_i$ \ are new non-minimal parameters, which may be safely
introduced for the massive regulators. The operators (\ref{op two
i}) act in the space of Dirac fermions, but may have either
fermionic or bosonic statistics. The corresponding multiplicities
are \ $d_i=(1,-2)$, like in the scalar case.
\beq
\bar{\Gamma}_{reg} &=&
-\frac12\,\Tr\log\left(\hat{H}\cdot\hat{H}_2^{*}\right)
\nonumber
\\
&+&  \frac{1}{2}\,\sum_{i=1}^N\,(-1)^{d_i}
\Tr\log\,\left(\hat{H}\cdot\hat{H}_2^{*}\right)_{(i)}\,.
\label{sum}
\eeq
The Pauli-Villars conditions for \ $s_i$ \ and \
$\mu_i$ \ are similar to (\ref{23a})-(\ref{23c}),
the only difference is the
sign of the contribution of the massless physical field, which is
opposite compared to the scalar case
\beq
\sum_{i=1}^N \, {s_i}
&=& -\,s_0\,=\,+\,1\,; \nonumber
\\
\sum_{i=1}^N s_i \mu_i^2   &=& \sum_{i=1}^N s_i\chi_i
= \sum_{i=1}^N s_i \mu_i^4  = \sum_{i=1}^N
s_i\chi_i^2  = 0
\label{plog f}
\eeq
and also we have here
factors $\,\chi_i\,$ instead of $\,\xi_i-1/6$.

The solution of equations (\ref{plog f}) is opposite in sign to
the one of (\ref{solutions mu xi}).
\beq
s_1 &=& -1,\qquad s_2=-4,\qquad
s_3=-s_4 =s_5=2,
\nonumber
\\
\mu_1^2 &=& 4,\quad \mu_2^2=3, \quad
\mu_3^2=1,\quad \mu_4^2=3,\quad \mu_5^2=4
\nonumber
\\
&\mbox{and}& \qquad
\chi_i=\mu^2_i\,.
\label{solutions mu xi f}
\eeq
Exactly as in the scalar case, another solution emerges
if taking the same \ $s_i,\,\mu_i^2$ \ and \ $\chi_i\equiv 0$.

Finally, we arrive, in the massless limit \ $m\to 0$, at the
following expression for the conformal anomaly
\beq
 <{\cal T}\!\!>  &=& \frac{1}{(4\pi)^2}\,\Big[
\,\frac{1}{20}\,C^2 - \frac{11}{360}\,E + \Big(\frac{1}{30} \!- \!
6\de\Big)\,{\square} R
\nonumber
\\
&+& 2h^2 \nabla_\al\phi\nabla^\al\phi - 2h^4\phi^4 +
\frac{h^2}{3}\, R\phi^2
\nonumber
\\
&-& \Big( \frac{2}{3}+\sigma\Big)\,h^2\square\phi^2\Big]\,,
\label{trace1112}
\eeq
where $\sigma$ and $\delta$  are parameters depending on details
of the Pauli-Villars regulators. This dependence indicates the
ambiguities of the coefficients of the total derivative terms \
$\square \phi^2$ \ and \ $\square R$. Qualitatively, the ambiguity
is the same as in the scalar case.


\section{\label{7}Conclusions}

We have investigated a problem of conformal anomaly and the
corresponding ambiguity in the theory of quantized matter fields
on the background which consists of the space-time metric and  an
additional scalar field. The anomaly can be evaluated via
different methods and perhaps the most simple one is based on the
dimensional regularization. However, the renormalization schemes
based on this regularization allow a much larger ambiguity in the
anomalous part of effective action. A similar ambiguity can be
also observed in covariant Pauli-Villars regularizations. Except
the specially designed artificial renormalization scheme described
in the Appendix, the ambiguity concerns only the local sector of
the effective action, while the non-local part of it is well
defined. In this respect our results are quite similar to the ones
obtained earlier in \cite{AGS} for the purely metric background
case. Although there is a serious difference between the two
cases. The ambiguity in the pure metric case concerns only the
vacuum sector, that is the effective action of external metric
field. In the present case of a metric-scalar background the
ambiguity concerns also the action of quantized scalar field.
Hence the effect of this ambiguity is much stronger. If one
attempts to fix the problem by introducing the special
renormalization condition, the theory becomes conformally
non-invariant already at the classical level. In this respect the
theory of conformal scalar field is closer to  quantum conformal
gravity \cite{Weyl}, rather than to the theory of free conformal
matter on purely metric background.

Let us notice that the renormalization scheme ambiguities are
common in the higher loop corrections to the $\beta$-functions
and therefore to the trace anomaly. For example, they have
been discussed in \cite{Jack,Osborn} for the case of quantum
field theory in curved space-time. The fundamental difference
between our results and the ones of the mentioned papers is
that we have found the origin of the trace anomaly ambiguity
which shows up already at the one-loop level. Furthermore, we
have clarified the existing mismatch  between the anomalous
violations of global and local conformal symmetries and found
that the ambiguities can arise not only within dimensional
regularization, but also for general covariant Pauli-Villars
regularizations.

Finally, the main lesson we learned from exploring the ambiguity
in the quantum violation of local conformal symmetry is that this
symmetry can not be exact even at the classical level. The model
with local conformal symmetry does not enable one to obtain a
consistent theory at quantum level, because the quantum
corrections are plagued by ambiguities. Our consideration also
leaves open the possibility to regard the conformal symmetry as an
approximation. In this case the conformal anomaly must be seen as
a useful way to evaluate the effective action and the
corresponding ambiguities may be easily fixed by implementing the
corresponding renormalization conditions on the coefficients of
local non-conformal terms in the renormalized action of the
theory. The approximate character of conformal symmetry will
manifest itself in a hierarchy of the parameters of the starting
action. The parameters in the non-conformal sector must be some
orders of magnitude smaller than the ones in the conformal sector.
In this case local non-conformal terms will be indeed modified by
the quantum (mainly anomalous) contributions, while the
ambiguities can be easily fixed through the renormalization
conditions in the non-conformal corner of the theory.

\vskip 2mm
\noindent
{\large\bf Acknowledgments.}
The work of the authors has been partially supported by the
 Spanish CICYT (grant FPA2004-02948) and DGIID-DGA
 (grant2005-E24/2) (M.A.), by research grants from CNPq (Brazil)
 and FAPEMIG (Minas Gerais, Brazil) (G.B.P. and I.Sh.),
 by the post-doctoral PRODOC fellowship from CAPES (G.B.P.)
 and by the long-term
 research fellowships from CNPq and ICTP (I.Sh.). I.Sh. is
 thankful to the Departamento de F\'{\i}sica Te\'orica at the
 University of Zaragoza for support and kind hospitality.


\appendix
\section{\label{A}Remark about more general
ambiguities in dimensional regularization}

The arbitrariness described in section 2 concerns only the local
finite term $\int d^4x\sqrt{g}R\phi^2$ in the effective action and
corresponding $\Box \phi^2$ term in the anomalous trace $<{\cal
T}>$. On the other hand, the presence of scalar field with a
non-trivial transformation law enables one to design
renormalizations schemes which make the arbitrariness in the
anomaly make the arbitrariness in the anomaly very much larger
than that derived via dimensional regularization in a usual way.
Let us notice that the conditions for the counterterms formulated
in section 2 (locality plus cancellation of the divergent part of
the effective action) can be satisfied in a great variety of
different ways in the presence of a scalar field. For example, in
the scalar field sector one can keep the transformation of scalar
field the same (\ref{conformal}), independent on the dimension
$n$. Then the derivation of anomaly in this sector will be
essentially the same as for the gravitational sector and the
anomaly $<{\cal T}>$ will be just proportional to the divergent
part of effective action, plus a standard $\Box \phi^2$-type
arbitrariness described above. On the other hand, in the purely
gravitational field one can use scalar field to eliminate the
anomaly completely. Let us choose the counterterm for the Weyl
term in the form \beq \De S_{Weyl}=\frac{\be_1}{n-4}\int
d^nx\sqrt{g} \,\phi^{n-4}\,C^2\,, \label{strange idea} \eeq where
we took the simplest transformation law (\ref{conformal}) of the
scalar field for simplicity\footnote{Choosing another version of
the transformation law in $n\neq 4$ one can always adjust the
power in (\ref{strange idea}) such that the result for the anomaly
would be the same.}. It is obvious that this counterterm is
conformal invariant and therefore using the standard procedure of
deriving anomaly from \cite{duff77} does not produce anomaly et
all. The same is true also for other anomalous terms. Therefore,
the combination of the dimensional regularization and scalar field
makes the whole anomaly arbitrary, not only the part corresponding
to the local sector of the one-loop effective action.

There are several reasons why such an ambiguous renormalization
scheme is unphysical. The dependence $\,\phi^{n-4}\,$ is
non-analytic on  the scalar field $\phi$ for non-integer values of
$n$ and for such a reason it can never be generated by radiative
perturvative corrections. Indeed, in perturbation theory Feynman
graphs have a definite integer number of external legs ($\phi's$)
which can not be a fractional number (see, however, Ref.
\cite{kogan} for scenarios with emergence of non-perturbative
non-analiticities). On the other hand the form of the counterterms
does not need to follow the form of the original quantum
corrections, but only satisfy the requirements formulated above.
Nevertheless, the scheme based on (\ref{strange idea}) is very
artificial and that the resulting ``total'' ambiguity does not
correspond to the direct calculations of a finite part of the
effective action. In general, the existence of the scheme
described above just shows that the dimensional regularization is
not an appropriate instrument to derive the conformal anomaly,
especially if the scalar field is present.
\vskip 12mm

\vskip 20mm


\end{document}